# Transmission electron microscopy of organic-inorganic hybrid perovskites: myths and truths

Shulin Chen[a, b], Ying Zhang[c], Jinjin Zhao[c]*, Zhou Mi[c], Jingmin Zhang[a], Jian Cao[b], Jicai Feng[b], Guanglei Zhang[c], Junlei Qi[b]*, Jiangyu Li[d]*, Peng Gao[a, e, f]*

[a]Electron Microscopy Laboratory, School of Physics, Peking University, Beijing 100871, China
[b]State Key Laboratory of Advanced Welding and Joining, Harbin Institute of Technology, Harbin 150001, China
[c]School of Materials Science and Engineering, School of mechanical Engineering, Shijiazhuang Tiedao University, Shijiazhuang, 050043, China
[d]Shenzhen Key Laboratory of Nanobiomechanics, Shenzhen Institutes of Advanced Technology, Chinese Academy of Sciences, Shenzhen 518055, Guangdong, China
[e]Collaborative Innovation Center of Quantum Matter, Beijing 100871, China
[f]International Center for Quantum Materials, School of Physics, Peking University, Beijing 100871, China

*Correspondence and requests for materials should be addressed to p-gao@pku.edu.cn (P. Gao); jy.li1@siat.ac.cn (J. Y. Li); jlqi@hit.edu.cn (J. L. Qi); jinjinzhao2012@163.com (J. J. Zhao)

**Abstract:** Organic-inorganic hybrid perovskites (OIHPs) have attracted extensive research interest as a promising candidate for efficient and inexpensive solar cells. Transmission electron microscopy characterizations that can benefit the fundamental understanding and the degradation mechanism are widely used for these materials. However, their sensitivity to the electron beam illumination and hence structural instabilities usually prevent us from obtaining the intrinsic information or even lead to significant artifacts. Here, we systematacially investigate the structural degradation behaviors under different experimental factors to reveal the optimized conditions for TEM characterizations of OIHPs by using low-dose electron diffraction and imaging techniques. We find that a low temperature (−180 ℃) does not slow down the beam damage but instead induces a rapid amorphization for OIHPs. Moreover, a less severe damage is observed at a higher accelerating voltage. The beam-sensitivity is found to be facet-dependent that a (100) exposed $MAPbI_3$ surface is more stable than (001) surface. With these guidance, we successfully acquire the atomic structure of pristine $MAPbI_3$ and identify the characterization window that is very narrow. These findings are helpful to guide future electron microscopy characterization of these beam-sensitive materials, which are also useful for finding strategies to improve the stability and the performance of the perovskite solar cells.

## 1. Introduction

Organic-inorganic hybrid perovskites (OIHPs) have achieved impressive improvements as a promising photovoltaic material, whose power conversion efficiency rapidly increases from initial 3.8% [1] to most recent 25.2% [2]. However the commercialization of the technology is still hindered by the poor long-term stability issues [3, 4]. Transmission electron microscopy (TEM)-based studies benefit the fundamental understanding of their nature, functionality as well as the degradation mechanism for these fascinating solar cell materials [5, 6], which makes great contribution to the development OIHPs based solar cells. OIHPs are extremely sensitive to the electron beam [7, 8] and it is reported that the high energy electron beam at 300 kV has already caused damage within 100 e $Å^{-2}$ and induced the degradation from $MAPbI_3$ to $PbI_2$ within 2200 e $Å^{-2}$ [7]. However, many studies ignored such electron beam sensitivity. Thus, even for basic phase identifications by relatively low dose electron diffraction (ED) technique, many researchers mistakenly identified the $PbI_2$ as $MAPbI_3$ [9-16]. The situation for in situ TEM studies on OIHPs is even worse since under common high resolution TEM (HRTEM) (Dose rate: thousands of e $Å^{-2}$ $s^{-1}$) or high-angle annular dark field scanning TEM (STEM) (Dose rate: hundreds or thousands of e $Å^{-2}$ $s^{-1}$) imaging modes, the total doses within several seconds are large enough to induce damage or full decomposition, likely leading to inaccurate or even incorrect conclusions (See supporting information Table S1 for detailed discussion). For example, Segawa et al. [17] used in situ HRTEM patterns to record the microstructural changes of $MAPbI_3$ for 5 minutes. At such a high dose, the sample is likely $PbI_2$ rather than $MAPbI_3$ thus the observed structural changes are likely due mainly to the electron beam instead of temperature effect. Also, Divitini G. et al. observed the heat-induced structural and chemical changes by in situ heating [11], however both the sample preparation by focused ion beam and STEM imaging mode can cause huge damage or complete decomposition, as a result the ion migration under heating likely takes place in $PbI_2$ rather than $MAPbI_3$. In situ electrical biasing TEM by Jeangros Q. et al. [18] and Shin B. et al. [19] also ignored the beam sensitivity of OIHPs. Therefore, it is highly desirable to investigate and clarify the effect of the electron beam itself on the structure instability so that we can draw valid conclusions from the TEM characterizations, especially under external stimuli by in situ TEM.

So far, not too many literatures have been devoted to investigating the effect of the electron beam illumination on OIHPs in TEM [7, 20-22]. Rothmann et al. noticed the dose rate for these extremely beam-sensitive materials and acquired selected area electron diffraction (SAED) pattern from the intrinsic MAPbI$_3$ at a low dose rate 1 e Å$^{-2}$ s$^{-1}$ [22]. Chen et al. uncovered that the structure damage has already been induced with 100 e Å$^{-2}$ s$^{-1}$ and proposed a detailed decomposition pathway of single crystalline MAPbI$_3$ [7]. Recently, Alberti A. et al. unveiled a Pb-clusters related degradation mechanism for polycrystalline MAPbI$_3$ films [20]. Besides these studies based on the ED patterns, Zhang et al. acquired the structure of CH$_3$NH$_3$PbBr$_3$ (MAPbBr$_3$) with extremely low doses using direct-detection electron-counting camera [23]. These TEM studies showed that dozens of e Å$^{-2}$ doses are able to induce damage or even decomposition for OIHPs, preventing atomic scale investigation as well as in situ study under multiply stimuli. To minimize radiation damage effect, Li et al. retrieved the structure of MAPbI$_3$ by cryo-electron microscopy (Cryo-EM), highlighting the great importance of low temperature for beam sensitive material characterization [8]. However, this is inconsistent with Rothmann's study that a lower temperature causes a rapid amorphization [21]. Thus whether or not the low temperature is beneficial for electron microscopy characterizations of OIHPs is still unclear. Furthermore, what kind of factors influence the beam sensitivity and how to increase the total damage-free doses during characterizations have been rarely explored and thus motivate this study.

In this work, we study the external factors (temperature, accelerating voltage) and internal factors (exposed facets) on the structural instabilities of OIHPs under electron beam irradiation. It is found that a low temperature (−180 ℃) causes a rapaid crystal-amorphous transition within low doses (129 to 150 e Å$^{-2}$), suggesting that low temperature is not helpful in preventing the electron beam damage while a high temperature (90 ℃) does not change the degradation pathway observed at room temperature. The beam damage mechanism is identified to be radiolysis since a high voltage is beneficial to reduce the damage. Besides we reveal that the beam-sensitivity is facet-dependent, with a (100) exposed MAPbI$_3$ surface more stable than the (001) surface. We also acquire the atomic structure of MAPbI$_3$ at an extremely low dose. Our findings can guide future electron microscopy characterization of these beam-sensitive materials and also lay a foundation for the in-depth decomposition study under various stimuli by electron microscopy.

**2. Experimental**

2.1 $CH_3NH_3PbI_3$ single crystalline film fabrication

$PbI_2$ and $CH_3NH_3I$ (molar ratio 1:1) were dissolved in γ-butyrolactone (GBL) with the concentration of 1.3 mol L$^{-1}$, prior to stirring in 12 hours at the 70 °C. The $CH_3NH_3PbI_3$ precursor solution was obtained and filtered using polytetrafluoroethylene (PTFE) filter with 0.22 μm pore size. The Fluorine-doped tin oxide (FTO)/TiO$_2$ substrates [24] were face to face clamped together at fixed distance of 50–200 μm. The fixed FTO/TiO$_2$ substrates were vertically and partial soaked in a 10 ml MAPbI$_3$ precursor solution at 120 °C, and then the feeding MAPbI$_3$ precursor solution was added twice a day in the nitrogen glove box. The perovskite solution climbs along the pores of the mesoporous TiO$_2$ substrate from bottom to top and covers the entire substrate, and then crystallizes into a film due to the temperature difference, forming the single crystal film [7, 24]. After some days, the substrates with MAPbI$_3$ single crystal film were taken out, and then dried at 120 °C for 10 minutes in nitrogen.

2.2 $CH_3NH_3PbBr_3$ single crystal preparation

$PbBr_2$ and $CH_3NH_3Br$ were dissolved in N,N-Dimethylformamide (DMF) and stirred at 30 °C for 12 hours to obtain 1 mol L$^{-1}$ MAPbBr$_3$ precursor solution. The MAPbBr$_3$ precursor solution was purified by polytetrafluoroethylene (PTFE) filter with 0.22 μm pore size, and then heated to 95 °C in a 10 ml container in dark environment. The container can be taken out by daylight at 30-50 % humidity until the MAPbBr$_3$ single crystals were grown after one night.[25]

2.3 TEM samples preparation

To avoid side reactions, all TEM samples were prepared in an argon-filled glovebox. We firstly scratched samples from substrate and dispersed them into anhydrous ether. Then, the clear suspensions were deposited on holey carbon copper girds. We sealed the carbon copper girds with a plastic bag full of argon before transformed into the TEM column. The water concentration inside the plastic bag is below 0.1 ppm. The perovskite powders were exposed inside the plastic bag for about 5 minutes during the transport of perovskite sample.

2.4 Characterization

Powder XRD patterns were obtained on D8 Advance diffractometer using Cu Kα radiation (40 kV and 40 mA) with a scanning rate of 4° min$^{-1}$ for wide-angle test increment. The HRTEM and the SAED patterns were conducted at an aberration corrected FEI (Titan Cubed Themis G2) operated at 80 kV and 300 kV. The EDS was carried out at 300 kV, 10–20 pA, 0.5–1 kcps for 120 s. The cooling and heating experiments were performed at FEI Tecnai F20 at 200 kV by a liquid nitrogen side-entry specimen holder (Gatan 636). The temperature is stable at the expected value for 1 hours before turning on the illumination to record data. As for all SAED images, it takes about 30 s to get the SAED patterns since the sample comes into sight. The HRTEM image of the MAPbI$_3$ is acquired at a magnification of 71 k by Gatan K2 direct-detection camera in the electron-counting mode with the dose fractionation function. The simulated ED patterns were obtained by the Single Crystalmaker software.

**3. Results and discussion**

To study the effects of temperature on the beam sensitivity, we first examine the phase of MAPbI$_3$ at different temperature, which is reported to be orthorhombic phase below −111±2 ℃, tetragonal phase between −111±2 ℃ and 58±5 ℃ and cubic phase over 58±5 ℃ [26, 27], as shown in Figure 1a–c. The MAPbI$_3$ is grown to be a tetragonal phase (Figure S1),[24] whose SAED pattern (Figure 1e) matches with the simulated one (Figure 1h). At −180 ℃, the acquired the SAED pattern (Figure 1d) shows no characteristic superstucture diffraction spots of the orthorhombic phase, highlighted by the circle on the simulated ED pattern (Figure 1g), suggesting that a low temperature in vacuum will not cause the transtion from tetragonl to orthorhombic phase for the single crystal MAPbI$_3$, which is also observed in Diroll's study [28]. We also examine the phase at a high temperature and find the SAED pattern at 90 ℃ (Figure 1f) indicates either a [110] direction of cubic phase (Figure 1i) or a [100] direction of the tetragonal phase (Figure 1h), thus making us unable to indentify the specific phase. Since the obtained SAED patterns can match with the pristine MAPbI$_3$, it is concluded the structure of MAPbI$_3$ is not damaged under low and high temperature in vacuum.

Then we further invetigate the degradation pathway at −180 ℃, 25 ℃ and 90 ℃ to reveal the effect of temperature on the beam sensitivity as shown in Figure 2. At −180 ℃, the SAED pattern (Figure 2a) is identified to be a [001] zone axis of tegragonal MAPbI$_3$ with additional superstucture diffraction spots marked by the yellow circles, which are possibly caused by the ordered iodine vacancies [7]. With increased doses, the sharp diffraction reflections continuously disappear and finally change into an amorphous ring within 150 e Å$^{-2}$ (Figure 2b–d), indicating a crystal-amorphous transition. The amorphous HRTEM image is shown in Figure S2. Compariably, a crystal-crystal transition from MAPbI$_3$ to PbI$_2$ is observed at 25 ℃ (Figure 2 e–h), whose degradation pathway starts with the loss of ordered halogen ions, followed by the loss of remaining halogen and methylamine ions, leading to final decomposition into PbI$_2$, as we reported ealier [7]. From [100] direction of tetragonal MAPbI$_3$, we again observe such crystal-amorphous transition at −180 ℃ within 129 e Å$^{-2}$ and a crystal-crystal transition at 25 ℃ (Figure S3). When temperature increases to 90 ℃, MAPbI$_3$ can maintain its stucture within 38 e Å$^{-2}$ (Figure S4), and the observed degradation pathway (Figure 2 i–l) is consistent to that at 25 ℃, which is through an intermidate phase to the final PbI$_2$. Figure 2m presents the total doses to observe the appearance of superstructure, transformation into amorpous phase and PbI$_2$ at different tempearture. At −180 ℃, the doses for generating superstructure (below 30 e Å$^{-2}$) and crystal-amorphous transition (150 e Å$^{-2}$) are smaller than 35 e Å$^{-2}$ and 475 e Å$^{-2}$ (transformation into PbI$_2$) at 25 ℃ and 38 e Å$^{-2}$ and 523 e Å$^{-2}$ (transformation into PbI$_2$) at 90 ℃, suggesting MAPbI$_3$ is less stable at low temperature under electron beam iiradiation. In fact, the data from earlier cryo-EM study also showed the formation of superstructure at 7.6 e Å$^{-2}$ [8], although it was not dicussed, and their dose is too low to observe amorphous transition. Rothmann et al., on the other hand, observed crystal-amorphous transition in poly-crystalline MAPbI$_3$ film, while no superstructure has been observed [21]. The total dose for becoming amorphous (~820 e Å$^{-2}$) in their study is larger than that in our case (150 e Å$^{-2}$), likely due to the enhanced stability with the appearance of orthorhomibic phase in tetragonal phase [29]. Such amorphlization transition only occurs at low temperature. It also should be noted that although the dose for transforming into PbI$_2$ at 90 ℃ (523 e Å$^{-2}$) is slightly larger than at 25 ℃ (451 e Å$^{-2}$), cosidering the possible variations of sample conditions such as different thickness [10] and crystalline quality from one specimen to another, it does not necessarily suggest a higher stability for MAPbI$_3$ at 90 ℃.

To comfirm the generalizaiton of such crystal-amorphous transtion at −180 ℃, we also investigate the structure evolution of cubic MAPbBr$_3$ (Figure S5) at different temperature as shown in Figure S6. With increased dose, MAPbBr$_3$ gradually decomposes to form intermediate phase with superstructure reflections similar to MAPbI$_3$, and eventually decomposing into final PbBr$_2$ (Figure S6e–h). Comparably, at −180 ℃, the sharp diffraction spots disappear quickly during the contious electron beam irradiation and finaly become an amorphous ring within 81 e Å$^{-2}$ (Figure S6a–d), again indicating a crystal-amorphous transition, which has also been observed in pure inorganic halide perovskite CsPbBr$_3$ [30]. We have carried out the energy dispersive X-ray spectroscopy (EDS) experiment to determine the composition of MAPbBr$_3$ at −180 ℃ as shown in Figure S7, which suggests the formation of Pb and PbBr$_2$ with negligible signal of N and C. However, the composition may be different from initial amorphous material due to the large electron dose illumination during the EDS measurement. Therefore, for both tetragonal MAPbI$_3$ and cubic MAPbBr$_3$, low temperature can not suppress the beam damage instead casuing a rapid crystal-amorphous transition. This is because under the electron beam irradiation, many defects (interstitials and vacancies) can be generated, which are mobile or form other quasi-stable configurations, causing new kinds of order at room temperature [31]. However the atomic defects are frozen and much less mobile [32, 33] at low temperature, thus they are prone to be accumulated as clusters, further becoming amorphous [21].

Besides temperature, accelerating voltage is another important factor concerning about beam sensitivity during TEM characterization. The total doses ($D_t$) before the superstructure spots appear is used as a reference to determine the beam sensitivity. We acquire the $D_t$ at 80 kV and 300 kV from samples on the same one TEM grid. As shown in Figure 3, the $D_t$ at 300 kV (38–39 e Å$^{-2}$) is about 2–3 times larger than that at 80 kV (13–16 e Å$^{-2}$) as shown in Figure S8, which suggests electron microscopy characterizations of OIHPs at a high voltage is helpful to reduce the damage. The smaller damage at high voltage also indicates knock-on damage is not the main damage mechanism since a higher energy incident electron is expected to cause a severer knock-on damage [34, 35]. The data recorded at 200 kV (Figure S9) presents consistent conclusion that higher voltage brings in smaller damage. Moreover, lowering temperature should help reduce the damage for heating damage mechanism but instead it is observed a rapid crystal-amorphous transition. Thus the damage mechanism for MAPbI$_3$ is identified to be radiolysis, which is consistent with its semi-conduct nature [36].

We also find a facet-dependent electron beam sensitivity for MAPbI$_3$. Specifically, as shown in Figure 4, the $D_t$ for a (100) exposed plane ranges from 210–500 e Å$^{-2}$ which is about 10 times larger than a (001) exposed plane (30–41 e Å$^{-2}$) for MAPbI$_3$, obtained from Figure S10. This is because the migration barrier of iodine on (001) surface (0.32 eV) is smaller than that on (100) surface (0.45 eV), calculated by the first-principles study [37], suggesting an easier diffusion of iodine on (001) surface and relatively higher stability of (100) surface. In fact, the higher stability of (100) exposed MAPbI$_3$ surface is also consistent with Lv's study that (001) facet exhibited greater sensitivity and faster eorsion rate to water than the (100) facet [38].

We further study the effect of anion and compare the beam sensitivity of tetragonal MAPbI$_3$ and cubic MAPbBr$_3$. The $D_t$ for MAPbI$_3$ ranges from 30 to 41 e Å$^{-2}$ (Figure S10a–c) which is about half of MAPbBr$_3$ (63–113 e Å$^{-2}$), acquired from Figure S11, suggesting that MAPbBr$_3$ is more stable than MAPbI$_3$ under electron beam iiradiation. The result is consistent with the conclution that MAPbBr$_3$ is more thermally and chemically stable than MAPbI$_3$ [39, 40], further indicating it is reasonable to judge the stability by comparing the $D_t$ values.

OIHPs are extremely sensitive to electron beam and it is always difficult to acqiure the atomic stucture. Our findings suggest TEM characterizations shall be carried out at 300 kV and RT rather than a low voltage and a low temperature. With these guidance, we have also acquired the structure of MAPbI$_3$ at 300 kV and 25 ℃, as shown in Figure 5. At a low dose (3.1 e Å$^{-2}$), the HRTEM image (Figure 5a) is identified to be the pristine MAPbI$_3$ judging from the corresponding fast Fourier transform (FFT) (Figure 5b), which is consistent with the simulated ED pattern in Figure 5c. When the summed dose increases to 24.9 e Å$^{-2}$, many additional superstructure diffraction spots appear (Figure 5d,e), which is likely due to the ordered vacancies in MAPbI$_{2.5}$, whose simulated ED (Figure 5f) can match the FFT, as we reported previously [7]. In fact, a few dim additional diffration spots have already appeared even at 6.2 e Å$^{-2}$ (Figure S12), suggesting several e Å$^{-2}$ is able to induce phase transition or damage for MAPbI$_3$. Accordingly, on the one hand, extra attention should be paid to the dim additional superstructure spots that are easily ignored but indicating the phase transition when dealing with the atomic structure of MAPbI$_3$. On the other hand, extremely low dose must be used to acquire the structure of MAPbI$_3$ as well as its structure evolution.

## 4. Conclusion

Recently, the cryo-EM is widely believed to mitigate the electron beam damage, especially for organic materials, by lessening mass loss and the heating damage to a certain degree [41, 42]. Particularly, the study of Li et al. [8] has inspired extensive investigations of cryo-EM for characterizing OIHPs [43, 44]. Our study reveals that for OIHPs cryo-imaging cannot prevent the damage but causes a rapid crystal-amorphous transition, suggesting the cryo-EM is not suitable for TEM characterizations of OIHPs. Besides the temperature effect, the recognized damage mechanism suggests a smaller damage of OIHPs is expected at a higher voltage.

Moreover, our work might shed lights on improving the performance of perovskite solar cells. For example, we find that heating to 90 ℃ does not cause more severe degradation of MAPbI$_3$. Therefore, more attention can be paid to stablize the interface or other layers in peroveskite solar cells to improve the high temperature performance. In addition, the facet-dependent beam sensitivity suggests that (100) surface might be more stable than (001) facet, which can also guide facet engineering to imporve the stability and performance of perovskite solar cells (PSCs) by growing (100)-textured perovskite films.

In summary, by using low-dose electron diffraction imaging techniques, we have investigated the optimized condition for TEM charactrization of OIHPs. We have also quantified the threshold electron dose to acquire the pristine SAED as well as the high-resolution TEM image of MAPbI$_3$. These findings are helpful to guide the furture TEM charactrization. On the other hand, our work provides some valuable insights into understanding the degradation mechanism of OIHPs and can also be useful for improving the performance of PSCs.

**Conflict of Interest**

The authors declare no conflict of interest.

**Acknowledgements**

The work was financially supported by the National Key R&D Program of China [grant numbers 2016YFA0300804, 2016YFA0300903, and 2016YFA0201001]; National Natural Science Foundation of China [grant numbers 51672007, 11974023, 51575135, U1537206, 11772207, and 11327902]; Key Area R&D Program of Guangdong Province (2018B010109009); National Equipment Program of China [grant number ZDYZ2015-1], and "2011 Program" Peking-Tsinghua-IOP Collaborative Innovation Center of Quantum Matter., Natural Science Foundation of Hebei Province for distinguished young scholar (A2019210204), High Level Talent Support Project in Hebei (C201821), State Key Laboratory of Mechanics and Control of Mechanical Structures, Nanjing University of Aeronautics and Astronautics (Grant No. MCMS-E-0519G04), China Scholarship Council (CSC), Youth Top-notch Talents Supporting Plan of Hebei Province. The authors acknowledge Electron Microscopy Laboratory in Peking University for the use electron microscopes.

**Author contributions**

P.G., J.L. J.Q. and J.Z. conceived the idea and designed the experiments. S.C. performed TEM related experiments with the help of J.M. Zhang and analyzed TEM data under the direction of P.G.; Y.Z., and Z.Z., synthesized the MAPbI$_3$ single crystal films and MAPbBr$_3$ single crystals under the direction of J.Z.; Z.M. and G.L. Zhang performed the SEM and XRD; J.C., J.F., and J.Q. provided crystals. S.C., J.L. and P.G. wrote the manuscript and all authors participated in the discussion.

**Appendix A. Supplementary materials**

Supplementary materials to this article can be found online or from the author.

**Figures and captions:**

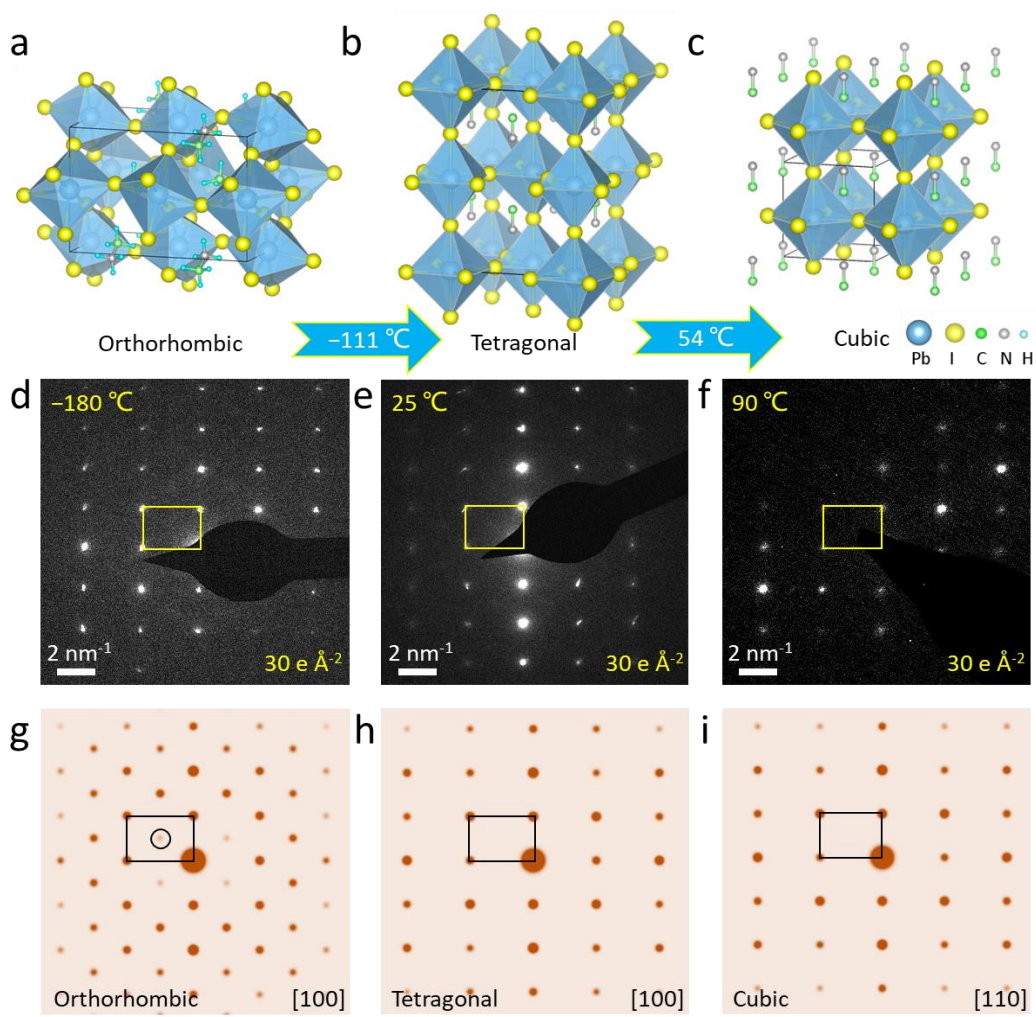

**Figure 1 Phases characterization of MAPbI$_3$ under different temperature.** Atomistic structures of a) orthorhombic (below −111 ℃), b) tetragonal (−111 to 54 ℃) and c) cubic (over 54℃) MAPbI$_3$. d, e, f) SAED patterns of MAPbI$_3$ at −180 ℃, 25 ℃, 90 ℃. g, h, i) The corresponding simulated SAED patterns along the [100] direction of orthorhombic MAPbI$_3$, [100] direction of tetragonal MAPbI$_3$ and [110] direction of cubic MAPbI$_3$.

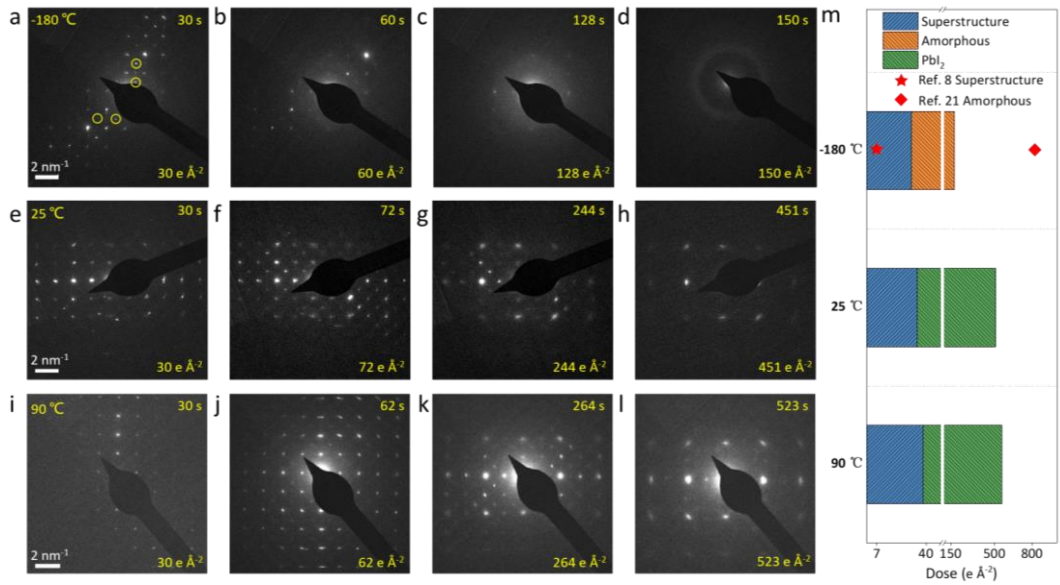

**Figure 2 The effect of temperature on the beam sensitivity of MAPbI$_3$.** Time-series SAED patterns along the [001] direction showing the degradation pathway a–d) at −180 ℃, e–h) at 25 ℃, i–l) at 90 ℃. The dose rate is 1 e Å$^{-2}$ s$^{-1}$ at 200 kV. m) The critical doses to observe the appearance of superstructure phase, transformation into amorphous phase and PbI$_2$. The doses marked by pentagram and rhombus are 7.6 e Å$^{-2}$ and 820 e Å$^{-2}$.

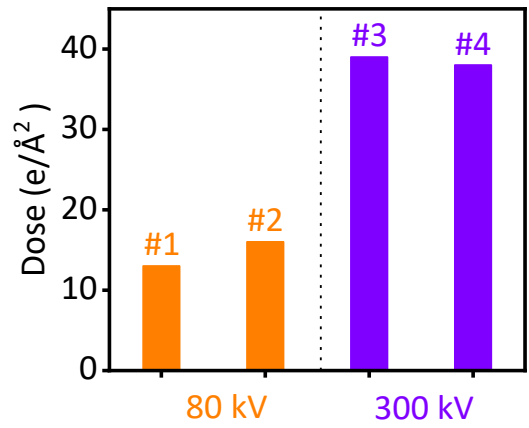

**Figure 3 The effect of voltage and electron beam damage mechanism.** The measured $D_t$ values at 80 kV and 300 kV of MAPbI$_3$, indicating a radiolysis mechanism that a high voltage can decrease the damage. The dose rate is 0.2 e Å$^{-2}$ s$^{-1}$ for #1–3 and 0.4 e Å$^{-2}$ s$^{-1}$ for #4).

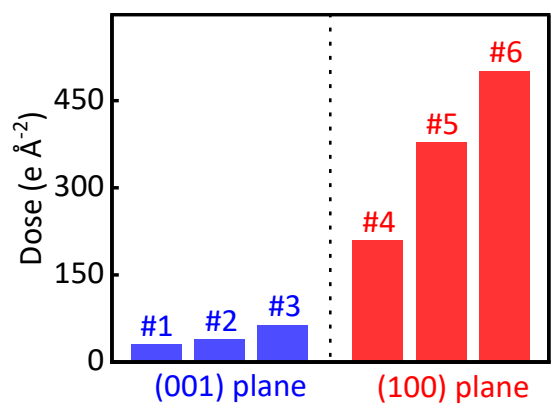

**Figure 4 Facet-dependent electron beam sensitivity.** The measured $D_t$ values with different crystal planes exposing of MAPbI$_3$. The dose rate is 1 e Å$^{-2}$ s$^{-1}$ at 300 kV.

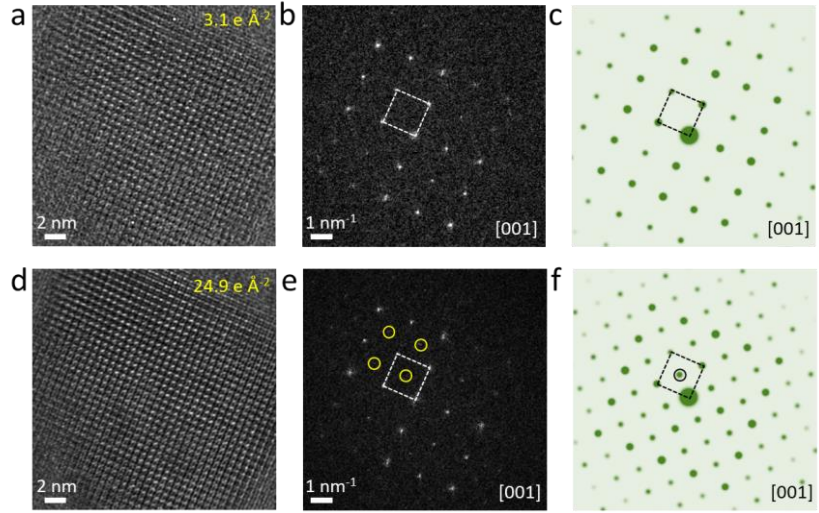

**Figure 5 Structure of MAPbI$_3$ and superstructure.** a) HRTEM image of MAPbI$_3$, b) the corresponding FFT pattern and c) the simulated ED pattern. d) HRTEM image of the superstructure, e) the corresponding FFT pattern and f) the simulated ED pattern. Circles highlight the additional superstructure spots.